\documentclass[reprint,amssymb, amsmath, aps, superscriptaddress, showpacs, footinbib, prb]{revtex4-1}
\usepackage{graphicx,epstopdf} 
\usepackage{amsmath}
\usepackage{tabu}
\usepackage{graphicx}

\newcommand{\be}{\begin{equation}}
\newcommand{\ee}{\end{equation}}
\newcommand{\bea}{\begin{eqnarray}}
\newcommand{\eea}{\end{eqnarray}}
\newcommand{\bse}{\begin{subequations}}
\newcommand{\ese}{\end{subequations}}

\begin{document}

\title{Structural and magnetic properties of spin-$1/2$ dimer compound Cu$_2$(IPA)$_2$(DMF)(H$_2$O) with a large spin gap}

\author{S. Thamban}
\affiliation{School of Physics, Indian Institute of Science Education and Research Thiruvananthapuram-695016, India}

\author{U. Arjun}
\affiliation{School of Physics, Indian Institute of Science Education and Research Thiruvananthapuram-695016, India}

\author{M. Padmanabhan}
\affiliation{Department of Chemistry, Amrita School of Arts and Sciences, Amrita Vishwa Vidyapeetham, Amrita University, Amritapuri Campus, Kollam-690525, India}

\author{R. Nath}
\email{rnath@iisertvm.ac.in}
\affiliation{School of Physics, Indian Institute of Science Education and Research Thiruvananthapuram-695016, India}

\begin{abstract}	
We present the synthesis and a detailed investigation of structural and magnetic properties of metal-organic compound Cu$_2$(IPA)$_2$(DMF)(H$_2$O) by means of x-ray diffraction, magnetization, and heat capacity measurements. Single crystals of the title compound were synthesized by judicious selection of organic ligand and employing a selective hydrothermal reaction route. It crystallizes in an orthorhombic structure with space group $Cmca$.
The structural analysis revealed that two Cu$^{2+}$ ions are held together by the organic component (-O-C-O-) in a square paddle-wheel to form spin dimers which are aligned perpendicular to each other and are further coupled through organic ligands (isophthalic acid) forming two-dimensional layers.
Temperature dependent magnetic susceptibility $\chi(T)$ could be described well using spin-$1/2$ dimer model. The spin susceptibility $\chi_{\rm spin} (T)$ shows an exponential decrease in the low temperature region, below the broad maximum, confirming the singlet ground state with a large spin gap of $\Delta/k_{\rm B} \simeq 409$~K. The heat capacity $C_{\rm p}$ measured as a function of temperature also confirms the absence of magnetic long-range-order down to 2~K.
\end{abstract}

\maketitle

\section{Introduction}
The last few decades have witnessed enormous research in the field of low-dimensional quantum magnets.\cite{Lemmens1} Model compounds especially with a gap in the excitation spectrum are extensively pursued. Spin gap is found in many low-dimensional quantum spin systems. Some examples include spin dimers,\cite{Sebastian617,Ruegg62,Koo263,Yogesh012407} integer spin chains,\cite{ShimizuR9835} even leg spin ladders,\cite{Azuma3463} alternating spin chains,\cite{Johnston219, Ahmed01650, Ghoshray214401, Tsirlin144412} spin-Peierls compounds,\cite{Hase3651} etc. Spin dimers are the simplest magnetic clusters, comprising of unified pair of magnetic ions, coupled antiferromagnetically. Such systems are characterized by a gap in their excitation spectrum and the energy difference between excited triplet states ($ S = 1 $) and singlet ground state ($ S = 0 $) constitutes the spin gap.

Recently, many new magnetic materials composed of antiferromagnetic spin dimers have been discovered. Because of different inter-dimer exchange networks, a variety of magnetic excitations have been observed under external magnetic field. Bose-Einstein-Condensation (BEC) of magnons in BaCuSi$_2$O$_6$, TlCuCl$_3$, (Ba,Sr)$_3$Cr$_2$O$_8$, NiCl$_2$-4SC(NH$_2$)$_2$ etc\cite{Sebastian617, Aczel207203, Aczel100409, Zapf077204, Ruegg62}, magnetization plateaus in Ba$_3$Mn$_2$O$_8$,[Ref.~\onlinecite{Uchida054429}] and magnetization plateaus and Wigner crystallization of magnons in SrCu$_2$(BO$_3$)$_2$[Ref.~\onlinecite{Kodama395,*Kageyama3168}] are the best known examples. These discoveries have accelerated the search for new spin dimer compounds and interesting quantum phenomena.

Unlike the inorganic materials (as discussed above), the metal-organic compounds are not explored well despite being easier to synthesize.
In magnetically active metal-organic systems, the organic moieties bond the metal ions in a wide variety of forms leading to structurally and magnetically diverse systems.\cite{Jain291,Carlin1986}
We have recently demonstrated that the organic moieties can modulate both the structural and magnetic features between metal ions significantly as the organic ligand NIPA leads to the formation of a hourglass nanomagnet Cu$_5$(NIPA)$_4$(OH)$_2$.11H$_2$O with two antiferromagnetic (AFM) and one ferromagnetic (FM) couplings between Cu$^{2+}$ ions\cite{Nath214417} while another similar ligand leads to the formation of a quasi-two-dimensional (2D) AFM with weakly anisotropic spin-$1/2$ square lattice.\cite{Nath054409}
Cu(1,3-$bdc$) is another well studied compound where Cu$^{2+}$ ions are connected via IPA forming a perfect Kagom$\acute{e}$ lattice with strong magnetic frustration.\cite{Nytko2922,*Chisnell214403,*Marcipar132402} It undergoes a magnetic long-range-order (LRO) at $T_{\rm N} \simeq 1.77$~K. We have reported that in such Kagom$\acute{e}$ lattices by changing the ligand it is possible to tune the magnetic ground state keeping the crystal structure intact.\cite{Ajeesh16}
Thus, the organic ligands play a vital role in stabilizing different ground states in the metal-organic complexes.

Herein, we report synthesis, crystal structure, and magnetic properties of a new spin-$1/2$ compound Cu$_2$(IPA)$_2$(DMF)(H$_2$O) [IPA = isophthalic acid and DMF = dimethyl formamide]. Cu$^{2+}$ ions in the compound form spin dimers which are weakly coupled via long ligands. Our magnetic susceptibility measurements revealed an activated behavior at low temperatures with a large spin gap of $\Delta/k_{\rm B} \simeq 409$~K.

\begin{figure}[t]
	\includegraphics[scale=0.24]{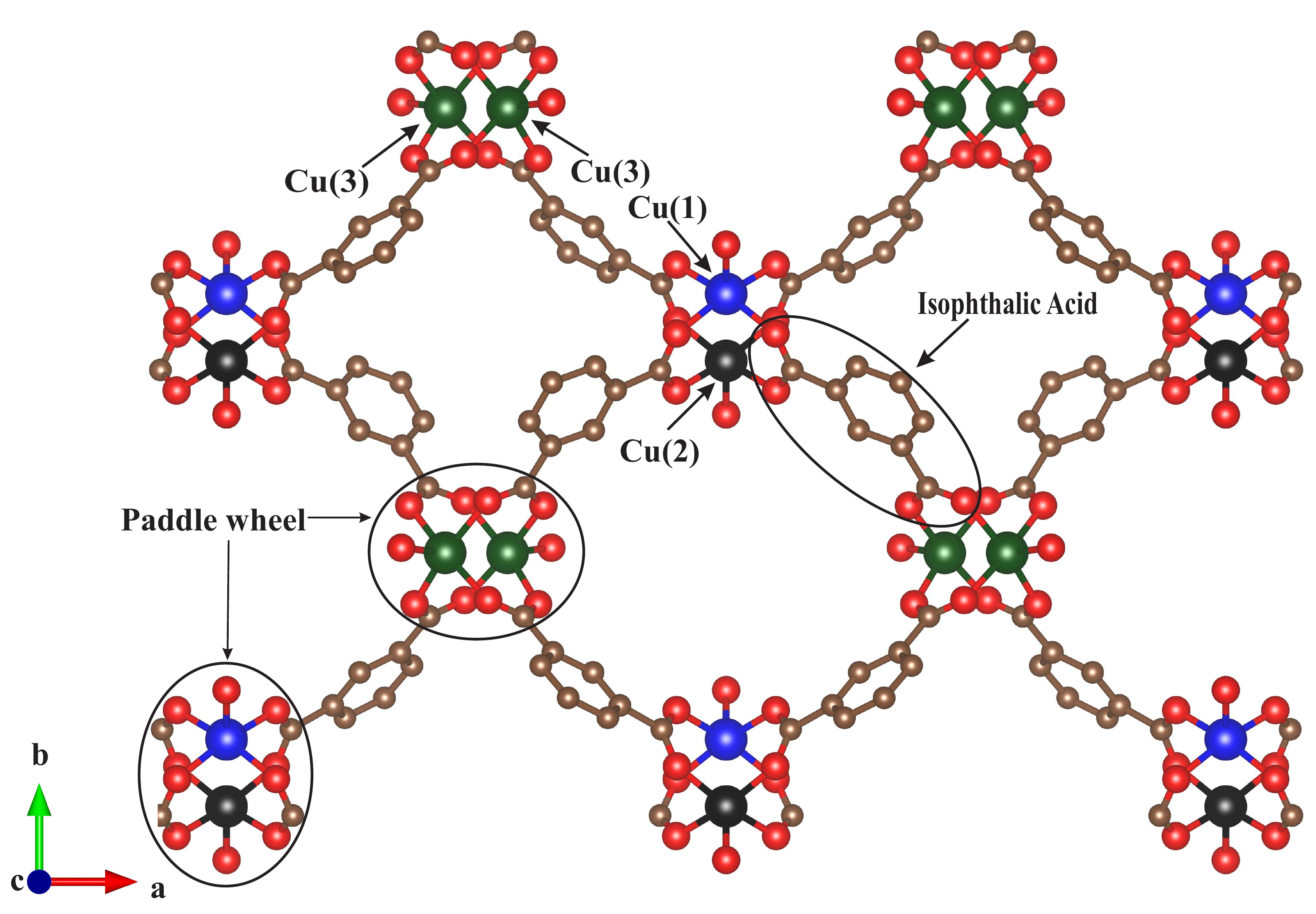}\\
	\caption{Projection of the two-dimensional layer on the $ab$-plane showing interconnecting and orthogonal network of Cu$^{2+}$ dimers in the crystal structure of Cu$_{2}$(IPA)$_{2}$(DMF)(H$_{2}$O). Since the DMF and H$_{2}$O molecules are not involved in the interaction path, we have removed them for clarity.}
	\label{Fig1}
\end{figure}
\section{\textbf{Experimental details}}
 All the reagent quality chemicals were obtained from Sigma-Aldrich and used without further purification. Single crystals of Cu$_{2}$(IPA)$_{2}$(DMF)(H$_{2}$O) were prepared by the hydrothermal route. In this process, an aqueous solution of Cu(NO$_{3}$)$_{2}$.3H$_{2}$O (0.5 mM, 0.12 g) was mixed with same equivalent of isopthalic (1,3-benzenedicarboxylic) acid (0.5 mM, 0.08 g) in 10 mL of DMF. The mixture was heated at 90 $^{0}$C in a 25 mL teflon capped autoclave. Big plate like blue crystals were collected from the walls of the vial after 3 days. The crystals after washing in water and acetone and then drying in air were found to be the phase pure form of Cu$_{2}$(IPA)$_{2}$(DMF)(H$_{2}$O). The yield was 72.6\% (based on Cu).

Single-crystal x-ray diffraction (XRD) was performed using a Bruker APEX-II diffractometer (MoK$_{\alpha1}$ radiation with wavelength $\lambda_{\rm avg} \simeq 0.71073$~\AA)  on a good-quality single crystal of Cu$_2$(IPA)$_2$(DMF)(H$_2$O) at room temperature and the crystal structure was solved.
The phase purity of crystals was also confirmed through powder XRD performed on the crushed powder sample at room temperature using a PANalytical (Cu$K_{\alpha}$ radiation, $\lambda_{\rm avg} \simeq 1.5418$~\AA) powder diffractometer. Magnetic susceptibility $\chi$ was measured as a function of temperature (2~K~$\leq T \leq$~380~K) using the vibrating sample magnetometer (VSM) attachment to the Physical Property Measurement System [PPMS, Quantum Design]. For high temperature measurements ($T\geq 380$~K), a high-$T$ oven (Model CM-C-VSM) was attached to the VSM. Our measurements were done upto 500~K, above which the sample was found to be decomposed. From the thermo gravity analysis, the decomposition temperature was indeed confirmed to be $\sim 500$~K.  Heat capacity, $C_{\rm p}(T)$ was also measured using heat capacity option of the PPMS on a small piece of sintered pellet, adopting relaxation technique.

\begin{figure}[h]
	\includegraphics[scale=0.35]{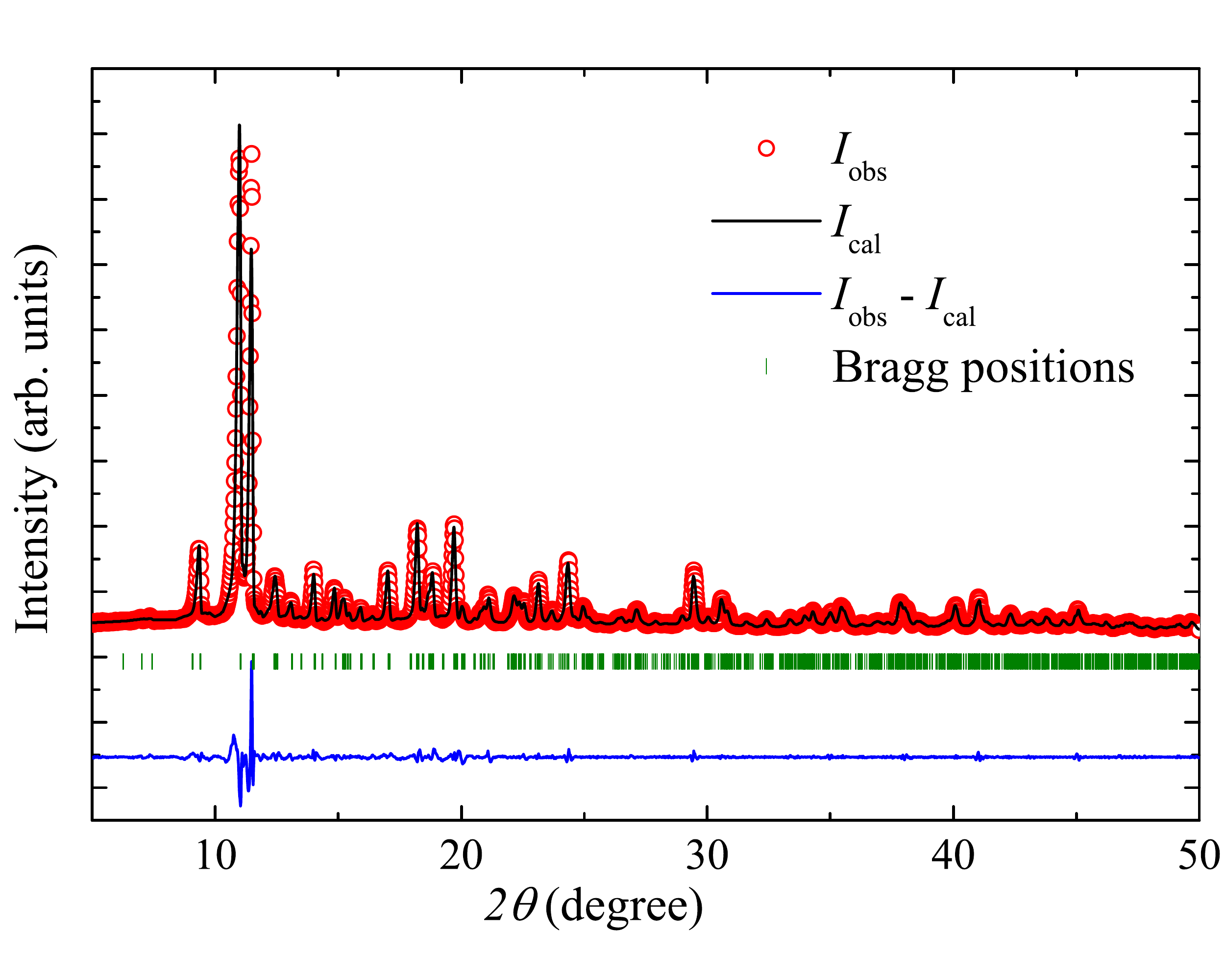}\\
	\caption{ Powder XRD pattern (open circles) at room temperature for Cu$_2$(IPA)$_2$(DMF)(H$_2$O). The solid line represents the Le-Bail fit, with the vertical bars showing the expected Bragg peak positions, and the lower solid line representing the difference between the observed and calculated intensities.}
	\label{Fig2}
\end{figure}
\section{\textbf{Results and Discussion}}
\subsection{\textbf{Crystallography}}
Crystal structure of Cu$_2$(IPA)$_2$(DMF)(H$_2$O)  was solved from single crystal XRD. It crystallizes in a orthorhombic structure with space group $Cmca$. The obtained structural parameters are tabulated in Table~\ref{structure} and ~\ref{Atomic}. Figure~\ref{Fig1} shows projection of the crystal structure in the $ab$-plane obtained from the single crystal XRD. It consists of paddle wheel like units, each containing two closely held Cu$^{2+}$ ions. In each paddle wheel, Cu$^{2+}$ ions are coupled via O-C-O path forming a spin dimer. It is to be noted that there are three inequivalent Cu sites [Cu(1), Cu(2), and Cu(3)] in the crystal structure which form two different dimers (type 1 and type 2). The intra-dimer distances are Cu(1)-Cu(2)~$\simeq 2.641$~\AA and Cu(3)-Cu(3)~$\simeq 2.633$~\AA for type 1 and type 2 dimers, respectively. Each dimer is connected to the adjacent dimers through the organic ligand (isophthalic acid, IPA) leading to either a coupled dimer or a quasi-2D layered structure in the $ab$-plane. The dimers are of course far apart ($\sim 8$~\AA) from each other where the inter-dimer interaction would be almost negligible compared to the intra-dimer interaction. Interestingly, in each layer the type 1 and type 2 dimers are aligned nearly perpendicular to each other. These 2D layers are well separated along the crystallographic $c$-direction.

 \begin{table}[htbp]
 	\setlength{\tabcolsep}{0.1cm}
 	\caption[Table1] {Crystal structure data of Cu$_{2}$(IPA)$_{2}$(DMF)(H$_{2}$O) obtained from single crystal XRD experiment at room temperature.}
 	\label{structure}
 \begin{tabular}{cc}
 		\hline\hline
 		Empirical Formula &  C$_{19}$H$_{15}$Cu$_{2}$NO$_{10}$ \\
 		Formula weight & 544.40 \\
 		Temperature & 296 K\\
 		Wavelength & 0.71073 \AA \\
 		Crystal system & Orthorhombic \\
 		Space Group &  $Cmca$ \\
    	Lattice parameters & $a = 28.059(3)$ \AA \\
 		                   & $ b = 25.025(2)$ \AA \\
   					 	   & $ c = 15.272(2)$ \AA  \\
						   & $ \alpha = 90^{0}$ \\
						   & $ \beta = 90^{0}$ \\
					 	   & $ \gamma = 90^{0}$ \\
 		Volume & 10724(2) \AA$^{3} $ \\
 		$Z$ & 16 \\
 		Calculated density $\rho_{cal}$ & 1.349 (mg/mm$^{3}$)\\
 		Absorption Coefficient & 1.631 mm$ ^{-1} $ \\
 		$F$(000) & 4384\\
 		Crystal size & $0.200 \times 0.200 \times 0.100 $ mm$^{3}$ \\
 		$ 2\theta $ ranges for data collection & 2.902$^{0}$ to 53.988$^{0}$ \\
 		Index ranges & $-35 \leq h \leq 35$\\
 		            & $-31 \leq k \leq 29$\\
 	            	& $-19 \leq l \leq 15$\\
 		Reflections Collected & 25810\\
 		Independent Reflections & 5964 [$R_{\rm int} = 0.0660$] \\
 		Data / restraints / parameters & 5964 / 0 / 295\\
 		Goodness of fit on $F^{2}$ & 0.946\\
 		Final $R$-Indices, $I \geq 2\sigma$(I) & $R _{1} \simeq 0.0429$, $wR _{2} \simeq 0.1102$ \\
 	    Final $R$-Indices(all data) & $R _{1} \simeq 0.0954$, $wR _{2} \simeq 0.1355$ \\
 		Largest diff. peak/hole & 0.650 \& -0.566 e \AA$^{-3}$\\
 		\hline\hline \\
 		
 	\end{tabular}
 \end{table}

\begin{table}[htbp]
	\setlength{\tabcolsep}{0.2cm}
	\caption{Atomic coordinates for Cu$_2$(IPA)$_2$(DMF)(H$_2$O). The isotropic atomic displacement parameters (ADP) $U_{\rm eq}$ are defined as one third of the trace of the orthogonal $U_{\rm ij}$ tensor. The errors are from the least-square structure refinement.}
	\label{Atomic}	
	\begin{tabular}{cccccc}
		\hline \hline
		&Atoms       &$x$     &$y$    &$z$      &$U_{\rm eq}$ \\
		&  & ($\times10^{4}$)&($\times10^{4}$) &($\times10^{4}$) &($\times10^{3}$ \AA$^2$) \\\hline
		&Cu1   & 5000 & 1150(1)  & 1820(1) &33(1) \\[0.1cm]
		&Cu2  &5000 &1904(1) &3030(1) &35(1)\\[0.1cm]
		&Cu3  &2187(1) &3997(1) &1857(1) &32(1)\\[0.1cm]
		&C1 &3667(1) &2656(1) &1630(3) &39(1)\\[0.1cm]
		&C2  &3937(1) &2284(1) &1171(3) &42(1)\\[0.1cm]	
		&C3  &3816(2) &2146(2) &321(3) &52(1)\\[0.1cm]
		&C4  &3419(2) &2370(2) &-68(3) &60(1)\\[0.1cm]
		&C5  &3145(2) &2732(2) &391(3) &52(1)\\[0.1cm]
		&C6  &3265(1) &2880(1) &1235(3) &39(1)\\[0.1cm]
		&C7  &2970(1) &3284(1) &1731(3) &35(1)\\[0.1cm]
		&C8  &4356(1) &2017(1) &1628(3) &39(1)\\[0.1cm]
		&C9  &1758(1) &5200(1) &3590(3) &37(1)\\[0.1cm]
		&C10 &1326(1) &5374(1) &3250(3) &37(1)\\[0.1cm]
		&C11  &6105(1) &826(1) &3613(3) &39(1)\\[0.1cm]	
		&C12  &6311(2) &1088(2) &4299(3) &56(1)\\[0.1cm]
		&C13  &6745(2) &910(2) &4630(3) &65(2)\\[0.1cm]
		&C14  &1970(2) &5473(2) &4279(3) &53(1)\\[0.1cm]
		&C15  &2015(1) &4731(1) &3186(3) &35(1)\\[0.1cm]
		&C16  &1359(2) &4438(2) &-839(4) &98(2)\\[0.1cm]
		&C17  &1960(3) &3857(3) &-1584(4) &118(2)\\[0.1cm]
		&C18  &1912(2) &3920(2) &7(4) &59(1)\\[0.1cm]
		&C19  &5662(1) &1049(1) &3197(3) &36(1)\\[0.1cm]
		&N1  &1753(2) &4067(2) &-779(3) &67(1)\\[0.1cm]
		&O1 &4507(1) &1601(1) &1281(2) &49(1)\\[0.1cm]
		&O2 &4504(1) &2231(1) &2309(2) &50(1)\\[0.1cm]
		&O3  &5495(1) &1459(1) &3559(2) &52(1)\\[0.1cm]
		&O4  &5500(1) &818(1) &2537(2) &45(1)\\[0.1cm]
		&O5 &1829(1) &4528(1) &2515(2) &45(1)\\[0.1cm]
		&O6 &2613(1) &4581(1) &1444(2) &44(1)\\[0.1cm]
		&O7  &2617(1) &3467(1) &1338(2) &43(1)\\[0.1cm]
		&O8  &3118(1) &3421(1) &2461(2) &45(1)\\[0.1cm]
		&O9 &1749(1) &4054(1) &710(2) &53(1)\\[0.1cm]
		&O10  &5000 &551(2) &804(4) &84(2)\\[0.1cm]
		&O11 &5000 &2456(2) &4123(3) &86(2)\\[0.1cm]
		&H1  &3754 &2754 &2195 &47\\[0.1cm]
		&H2 &4002 &1903 &13 &62\\[0.1cm]
		&H3	&3336 &2277	&-638 &72\\[0.1cm]
		&H4 &2875 &2879 &130	&62\\[0.1cm]
		&H5	&1184 &5193	&2786 &44\\[.01cm]
		&H6 &6161 &1384 & 4544 &68\\[0.1cm]
		&H7	&6885 &1091& 5097& 78\\[0.1cm]
		&H8 &2261 &5360 &4500 &64\\[0.1cm]				
		\hline
		\hline
	\end{tabular}
\end{table}

Le-Bail fit of the observed powder XRD pattern was performed using the \verb FullProf  package.\cite{Carvajal55} The initial structural parameters for this purpose were taken from the single crystal XRD data. Figure~\ref{Fig2} shows the powder XRD pattern of Cu$_2$(IPA)$_2$(DMF)(H$_2$O) at room temperature along with the calculated pattern. All the peaks could be fitted using the orthorhombic ($Cmca$) structure. The obtained best fit parameters are $a = 28.275(1)$~\AA, $b = 25.231(1)$~\AA, $c = 15.3701(5)$~\AA and the goodness-of-fit $\chi^2 \simeq 8.27 $. These lattice parameters are close to the values obtained from the single crystal XRD.

\subsection{\textbf{Magnetic Susceptibility}}
\begin{figure}[h]
	\includegraphics[scale=0.34]{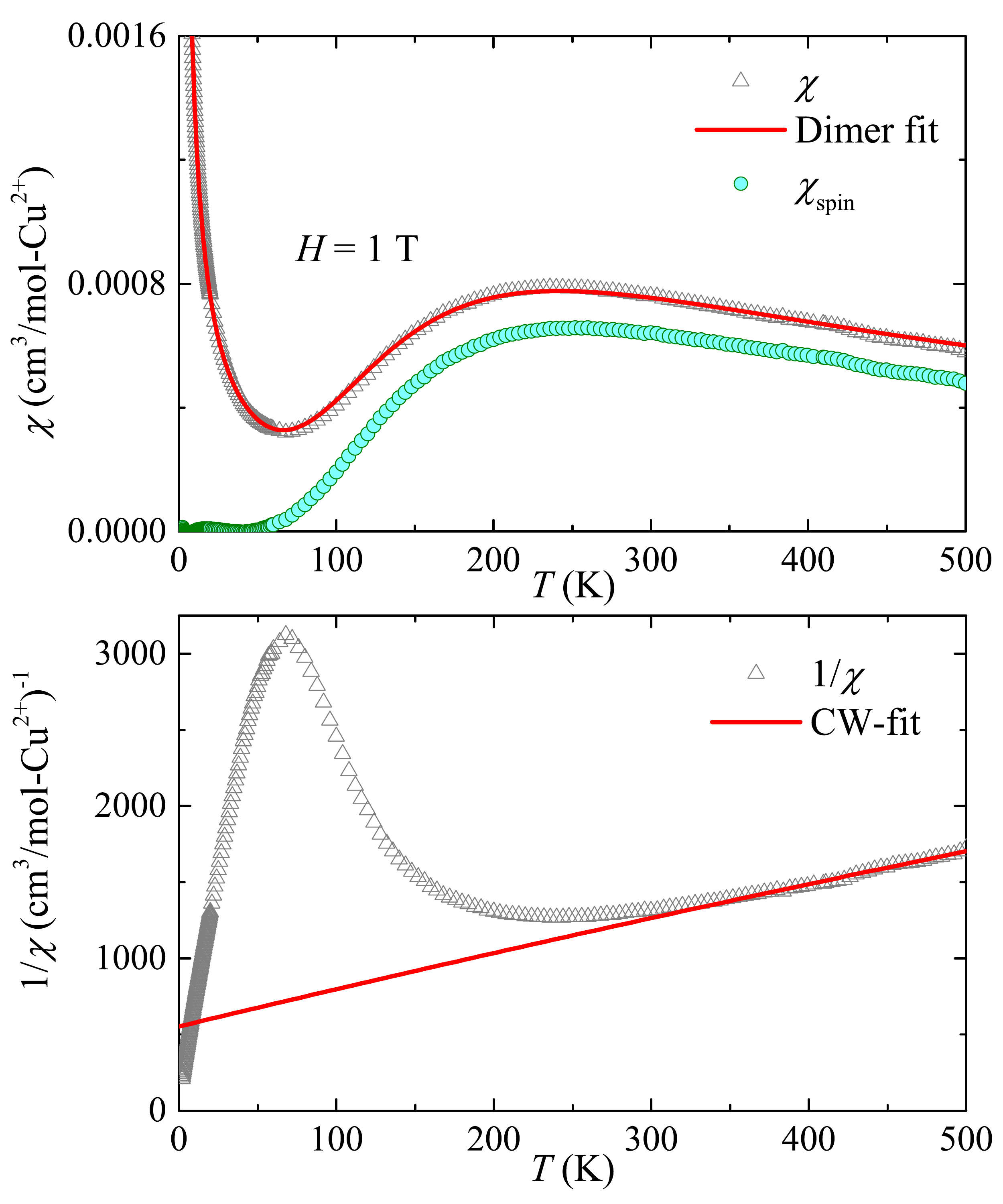}\\
	\caption{Upper panel: $\chi(T)$ measured at $H= 1$~T. The solid line represents the fit using Eq.~\eqref{chi_cw_dimer}. Lower panel: $1/\chi$ vs. $T$ and the solid line is the CW fit using Eq.~\eqref{cw}}
	\label{Fig3}
\end{figure}
Magnetic susceptibility $\chi(T)$ ($\equiv M/H$, where $M$ is the magnetization) measured in an applied field of $H=1$~T is shown in the upper panel of Fig.~\ref{Fig3}. As the temperature decreases, $\chi(T)$ increases and passes through a broad maximum near $T_\chi^ {\rm {max}}$ $\simeq200$~K, a short range magnetic order which is fingerprint of low dimensional antiferromagnetic spin systems. At very high temperatures ($T \gg$ exchange coupling, $J/k_{\rm B}$), usually spins are randomly oriented and $\chi(T)$ behaves like a paramagnet. However, in our compound $\chi(T)$ does not seem to have reached the paramagnetic region even at $500$~K. Thus, persistence of magnetic correlations upto such a high temperature clearly reflects a strong antiferromagnetic exchange coupling between Cu$^{2+}$ ions.  At low temperatures ($T\lesssim 70$~K), $\chi(T)$ shows a clear upturn, possibly due to extrinsic paramagnetic impurities and/or defects present in the sample. No signature of any magnetic long-range-order (LRO) is observed down to 2~K.

To extract the magnetic parameters, $\chi(T)$ at high temperatures was fitted by the following expression
\begin{equation}\label{cw}
\chi(T) = \chi_0 + \frac{C}{T + \theta_{\rm CW}},
\end{equation}
where $\chi_0$ is the temperature independent contribution consisting of core diamagnetic susceptibility ($\chi_{\rm core}$) of the core electron shells and Van-Vleck paramagnetic susceptibility ($\chi_{\rm VV}$) of the open shells of the Cu$^{2+}$ ions present in the sample. The second term in Eq.~\eqref{cw} is the Curie-Weiss (CW) law with the CW temperature ($\theta_{\rm CW}$) and Curie constant $C = N_{\rm A} \mu_{\rm eff}^2/3k_{\rm B}$, where $N_{\rm A}$ is Avogadro number, $k_{\rm B}$ is Boltzmann constant, $\mu_{\rm eff} = g\sqrt{S(S+1)}$$\mu_{\rm B}$ is the effective magnetic moment, $g$ is the Land$\acute{\rm e}$ $g$-factor, $\mu_{\rm B}$ is the Bohr magneton, and $S$ is the spin quantum number. As shown in the lower panel of Fig.~\ref{Fig3}, we fitted the $1/\chi(T)$ data in the temperature range 400~K to 500~K using Eq.~\eqref{cw}. Since the compound is not completely paramagnetic in this temperature range, we fixed the value of $C$ to 0.375~cm$^3$K/mol-Cu$^{2+}$ (expected $C$ value for spin-$1/2$ systems taking $g=2$ to give $\mu_{\rm eff} = 1.73$~$\mu_{\rm B}$/Cu$^{2+}$) and tried to vary other parameters. The fit yields $\chi_0 \simeq 7.94 \times 10^{-5}$~cm$^3$/mol-Cu$^{2+}$ and $\theta_{\rm CW} \simeq 204.84$~K. The positive value of $\theta_{\rm CW}$ indicates that the dominant exchange couplings between Cu$^{2+}$ ions are antiferromagnetic in nature. Such a large value of $\theta_{\rm CW}$ also suggests a strong exchange interaction between Cu$^{2+}$ ions.

In order to get an estimation of the exchange coupling and to have a better view of the spin-lattice, we fitted the observed $\chi(T)$ data to the following expression\cite{Johnston9558}
\begin{equation}\label{chi_cw_dimer}
\chi(T) = \chi_0 + \frac {C_{\rm imp}}{T+\theta_{\rm imp}}+\chi_{\rm dimer},
\end{equation}
where, $C_{\rm imp}$ represents the impurity concentration and $\theta_{\rm imp}$ provides the effective interaction strength between impurity spins. $\chi_{\rm dimer}$ is the expression for exact spin susceptibility of a spin-$1/2$ dimer model which can be written as\cite{Yogesh012407,Nakajima054706}
\begin{equation}\label{chi_dimer}
\chi_{\rm dimer} =\frac{N_{\rm A} g^2 \mu_{\rm B}^2}{k_{\rm B} T[3 + exp(\Delta/k_{\rm B}T)]}.
\end{equation}
The dimers have a spin gap $\Delta/k_{\rm B}$ ($=J/k_{\rm B}$) between the singlet ground state and the triplet excited states. Here, $J/k_{\rm B}$ is the intra-dimer exchange coupling.

The low temperature upturn of $\chi(T)$ below 70~K originates from the isolated Cu$^{2+}$ ions possibly due to lattice defect. The second term (CW term) is included in Eq.~\eqref{chi_cw_dimer} to account for this low temperature upturn. Overall, Eq.~\eqref{chi_cw_dimer} has six fitting parameters: $\chi_0$, $C_{\rm imp}$, $\theta_{\rm imp}$, $g$, and $J/k_{\rm B}$. As shown in the upper panel of Fig.~\ref{Fig3}, Eq.~\eqref{chi_cw_dimer} fits very well to the $\chi(T)$ data over the whole temperature range. The obtained best fit parameters are $\chi_0 \simeq 7.50 \times 10^{-5}$~cm$^3$/mol-Cu$^{2+}$, $C_{\rm imp}\simeq0.015$~cm$^3$K/mol-Cu$^{2+}$, $g\simeq1.91$, $\Delta/k_{\rm B} \simeq 409.83$~K, and $\theta_{\rm imp} \simeq 0.96$~K. This value of $C_{\rm imp}$ corresponds to a spin concentration of nearly 4.0~\%, assuming the impurity spins as spin-$1/2$.
In order to demonstrate the non-magnetic ground state, $\chi_0 + \frac {C_{\rm imp}}{T+\theta_{\rm imp}}$ was subtracted from the $\chi(T)$ data and the obtained intrinsic spin susceptibility $ \chi_{\rm spin} (T)$ is presented in the upper panel of Fig.~\ref{Fig3}. It is evident that $ \chi_{\rm spin} (T)$ decreases rapidly towards zero at low temperatures which unambiguously establishes a spin gap in the excitation spectrum.\cite{Tsirlin144412,Tsirlin104436}

It is to be noted that our attempt to fit the $\chi(T)$ data using coupled dimer model didnot improve the fitting significantly and the obtained value of inter-dimer coupling ($J^{'}$) was not at all reliable. This implies that the spin-lattice behaves more like isolated dimers rather than coupled dimers. Typically, in isolated dimers, the strength of intra-dimer coupling quantitatively reflects the magnitude of spin gap. Moreover, the value of spin gap can be reduced immensely by the inter-dimer coupling, upto a threshold value, above which there is a transition to magnetic LRO.\cite{Tsirlin144412,Arjun2016} Thus, such a large spin gap in Cu$_2$(IPA)$_2$(DMF)(H$_2$O) also implies a negligible inter-dimer coupling in the compound. This finding is in reasonable agreement with the structural data where the dimers are well separated ($\sim 8$~\AA) from each other and are connected via a long and extended path. As one can see from Fig.~\ref{Fig1} that Cu$^{2+}$ dimers are formed through a complex path of -O-C-O- and the arrangement of Cu$^{2+}$-O-C-O-Cu$^{2+}$ is not in one line. Usually, one expects a strong exchange interaction for the interaction path involving only O between metal ions and $\sim 180^{0}$ metal-oxygen-metal bond angle. Despite having a complex interaction path within the dimers, this compound shows a large spin gap of $\Delta/k_{\rm B} \simeq 409$~K which is a surprising feature of the compound. It is worth mentioning that spin dimer compounds such as TlCuCl$_3$, (Sr,Ba)$_{3}$Cr$_{2}$O$_{8}$ etc where the super-exchange involves only oxygen/chlorine atom between metal ions gives a rather small value of the spin gap.\cite{Yogesh012407,Nakajima054706,Ruegg62}

There are a few spin-$1/2$ metal-organic complexes based on Cu$^{2+}$ reported to have paddle-wheel like dimer units but with different ground states. For instance, [Cu$_{2}$(OOCC$_{5}$H$_{11}$)$_{4}$(urea)$_{2}$], 
[$\{ {\rm Cu}_{2}({\rm OOCC}_{5} H_{11})_{4}({\rm urea})\}_{2}$], [Cu$_{2}$(OOCC$_{5}$H$_{11}$)$_{4}$]$_n$, and Cu$_{2}$(O$_{2}$CCH=CHCH$_{3}$)$_{4}$(DMF)$_{2}$ show singlet ground state with a spin gap of $\sim342$~K, $\sim375$~K, $\sim333$~K, and $\sim220$~K, respectively.\cite{Kozlevvcar4220,Schlam88} The compound tetrakis($\mu_{2}$benzoato-O,O$^\prime$)-bis(dimethyl sulfoxide)dicopper(II) which also has paddle-wheel like dimer units but with a different ligand show one-dimensional character with a exchange coupling $J/k_{\rm B} \simeq 405$~K.\cite{Reyes16312} It undergoes a FM ordering at $T_{\rm C} \simeq 9$~K. Recently, another compound Cu(IPA)(H$_2$O) is reported to have paddle-wheel like spin dimers coupled through IPA ligand, similar to our compound. The crystal lattice forms a Kagom$\acute{e}$ structure with a very large spin gap of $\Delta/k_{\rm B} \simeq 410$~K.\cite{Mantas2016} From the above examples it is clear that though all these compounds are having paddle-wheel like spin dimers but different connecting ligands between dimers lead to diverse ground states. Thus, in these compounds one can tune the ground state properties and even the magnitude of the spin gap/exchange coupling by simply changing/modifying the connecting ligand.

\subsection{\textbf{Magnetic Isotherm}}
\begin{figure}[h]
	\includegraphics[scale=0.34]{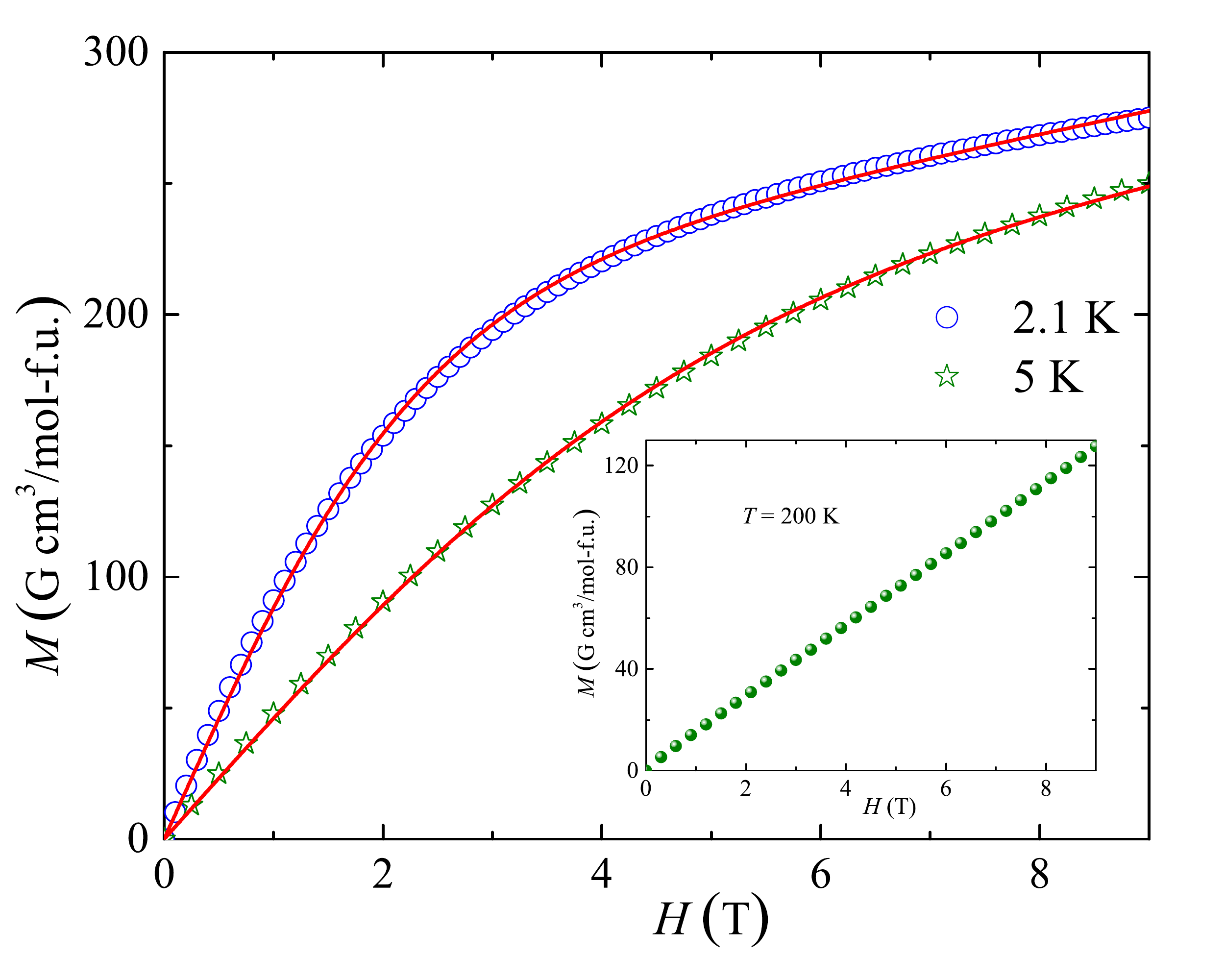}\\
	\caption{Magnetic isotherms ($M$ vs. $H$) at $T$= 2.1~K and 5~K. Solid lines are the fits using Eq.~\eqref{brill_eq}. Inset: $M$ vs. $H$ at $T=200$~K which shows linear behavior.}
	\label{Fig4}	
\end{figure}
As one can see in the upper panel of Fig.~\ref{Fig3}, the magnitude of $ \chi_{\rm spin} (T)$ is nearly zero below 50~K. Thus, the low temperature upturn observed in $\chi(T)$ below 50~K is purely extrinsic in nature. Therefore, one can precisely estimate the extrinsic contribution to $\chi(T)$ by analyzing the magnetic isotherms ($M$ vs $H$) at low temperatures. Figure~\ref{Fig4} shows the magnetic isotherms measured at different temperatures. At $T = 200$~K (inset of Fig.~\ref{Fig4}), it is a nice straight line. As the temperature is lowered, it shows a pronounced curvature typically observed for paramagnets. In order to quantitatively estimate the paramagnetic impurity contribution, we fitted the $ M $ vs $ H $ data at $T$ = 2.1~K and 5~K by the following equation\cite{Nath174513}
\begin{equation}\label{brill_eq}
M = \chi H + f_{\rm imp} N_{\rm A}  g_{\rm imp} \mu_{\rm B} S_{\rm imp} B_{S_{\rm imp}}(x),
\end{equation}
where $  \chi  $ is the intrinsic susceptibility of the sample, $f_{\rm imp}$ is the molar fraction of  the impurities, $N_{\rm A}$ is the Avogadro's number, $g_{\rm imp}$ impurity $g$-factor, $  S_{\rm imp} $ is the impurity spin, $ B_{S_{\rm imp}}(x)  $ is the Brillouin function,\cite{Kittel} and the modified argument of the Brillouin function is $x=  g_{\rm imp} \mu_{\rm B}  S_{\rm imp}H/[k_{\rm B}(T-\theta_{\rm imp})]$.

\begin{figure}[h]
	\includegraphics[scale=1]{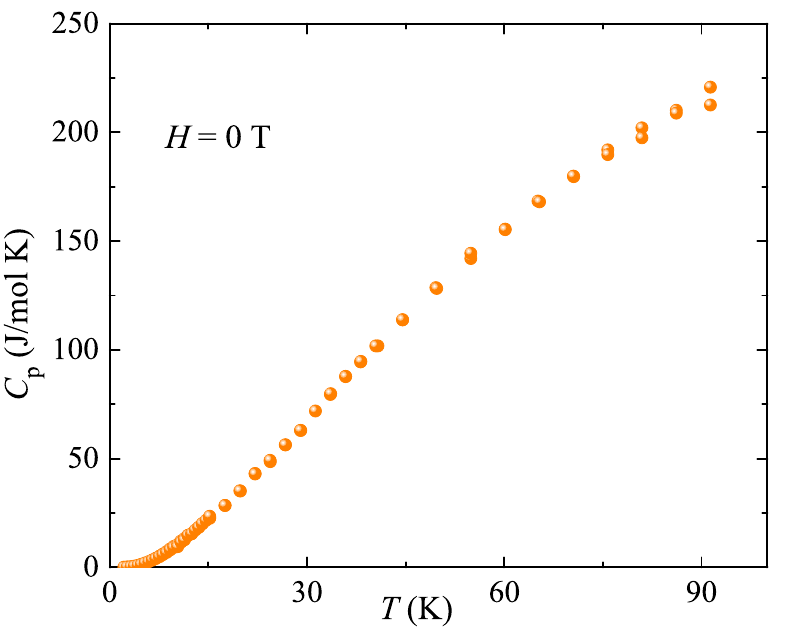}\\
	\caption{Temperature dependent heat capacity $C_{\rm p}(T)$ of Cu$_2$(IPA)$_2$(DMF)(H$_2$O) measured at zero applied field.}
	\label{Fig5}
\end{figure}

In order to reduce the number of fitting parameters, in the Brillouin function fit using Eq.~\eqref{brill_eq}, we fixed  $g_{\rm imp}$ and  $S_{\rm imp}$ to 1.9 [obtained from the $\chi(T)$ analysis] and $1/2$, respectively. The fitting parameters ($\chi$, $f_{\rm imp}$, and $\theta_{\rm imp}$) obtained at $T=2.1$~K and 5~K are tabulated in Table~\ref{Johnston_chi}. The value of $f_{\rm imp}$ obtained from the analysis of 2.1~K data was fixed while analysing the 5~K data. Using the value of $f_{\rm imp}$ in $C_{\rm imp} = f_{\rm imp} N_{\rm A} g_{\rm imp} \mu_{\rm B} S_{\rm imp} (S_{\rm imp}+1)$, we calculated the Curie constant of the paramagnetic impurities to be $C_{\rm imp} \approx 0.0127$~cm$^{3}$~K/mol. This value of $C_{\rm imp}$ corresponds to $\sim 3.4\%$ of spin-$1/2$ impurity spins which is close to that obtained from the $\chi(T)$ analysis.
\begin{table}[htbp]
	\setlength{\tabcolsep}{0.1cm}
	\caption{ Parameters obtained from the fitting of  $M$ vs. $H$ curves by the Brillouin function [Eq~\eqref{brill_eq}].}
	\label{Johnston_chi}
	\begin{tabular}{cccc}
		\hline \hline
		Temperature & $\chi$ & $f_{\rm imp}$ & $\theta_{\rm imp}$ \\
		(K)& (cm$^3$K/mol) & (mol\%) & (K) \\\hline
		2.1 & $\sim$8.7 $\times$ 10$^{-4}$ & $\sim$0.03759 & $\sim$0.59 \\
		5 & $\sim$6.6 $\times$ 10$^{-4}$& $\sim$0.03759 & $\sim$1.81\\
	
		\hline \hline
	\end{tabular}
\end{table}

\subsection{\textbf{Heat Capacity}}
The heat capacity $C_{\rm p}$ of Cu$_2$(IPA)$_2$(DMF)(H$_2$O) is measured as a function of temperature at zero applied field in the temperature range from 2~K to 90~K (see Fig.~\ref{Fig5}). No indication of magnetic LRO was observed down to 2~K, which is in accordance with our $\chi(T)$ data. In a magnetic insulator, $C_{\rm p}(T)$ has two major contributions: one due to phonon excitations and the other due to magnetic part. We refrained from doing any quantitative analysis of the $C_{\rm p}$ data since it was not possible to subtract the phonon contribution without a non-magnetic analogue compound.

\section{\textbf{Summary}}
We have synthesized a novel paddle-wheel type Cu$^{2+}$ spin-$1/2$ dimer compound Cu$_2$(IPA)$_2$(DMF)(H$_2$O) and investigated its crystal structure and magnetic properties. It crystallizes in an orthorhombic ($Cmca$) crystal structure. Analysis of $\chi(T)$ revealed dimer model for the spin lattice  with a large spin gap of $\Delta/k_{\rm B} \simeq 409$~K. The paramagnetic impurity concentration was estimated to be $\sim 4$\% which is likely due to defects present in the sample. No signature of magnetic LRO was observed down to 2~K from the $C_{\rm p}(T)$ data, further supporting the gapped behaviour. In magnetic insulators, strongest super-exchange interaction between magnetic ions is usually expected via non-magnetic ions such as oxygen.\cite{Motoyama3212} However, in our compound, despite having a long and complex interaction path Cu-O-C-O-Cu, the intra-dimer interaction is surprisingly very large.
For a comparison, the highest value of spin gap is found to be reported for SrCu$_2$O$_3$ ($\Delta/k_{\rm B} \simeq 420$~K)\cite{Azuma3463} among the inorganic compounds and for Cu(IPA)(H$_{2}$O) ($\Delta/k_{\rm B} \simeq 410$~K) among the metal-organic compounds.\cite{Schlam88} Thus, the value of spin gap found for Cu$_2$(IPA)$_2$(DMF)(H$_2$O) is nearly same in magnitude to the highest values reported so far.

As pointed out earlier, the crystal structure contains quasi-2D layers of orthogonal Cu$^{2+}$ dimers. Such a crystal lattice is reminiscent of the magnetic sub-lattice reported for SrCu$_2$(BO$_3$)$_2$ where Wigner crystallization of magnons and magnetization plateaus were observed in an applied magnetic field.\cite{Kodama395}
Thus, a large spin gap and orthogonal arrangement of dimers in the 2D layer make Cu$_2$(IPA)$_2$(DMF)(H$_2$O) a unique system for further investigations. Furthermore, the value of spin gap is very sensitive to the inter-dimer couplings which tend to reduce the spin gap and heavily affect the properties under high magnetic field. In such a context, this compound serves as a starting point to synthesize further new compounds where one can tune the spin gap simply by changing the organic ligand. 

\section{\textbf{Acknowledgement}}
We would like to acknowledge  IISER Thiruvananthapuram for providing necessary experimental facilities.

%

\end{document}